\documentstyle[12pt]{article}
\newcommand{\be}{\begin{equation}}
\newcommand{\ee}{\end{equation}}
\newcommand{\ben}{\begin{eqnarray}
\displaystyle}
\newcommand{\een}{\end{eqnarray}}

\newcommand{\la}{{\lambda}}
\newcommand{\Si}{{\Sigma}}
\newcommand{\de}{{\delta}}

\newcommand{\bde}{{\bar \delta}}

\newcommand{\cK}{{\cal K}}

\newcommand{\cJ}{{\cal J}}

\newcommand{\cH}{{\cal H}}

\newcommand{\na}{\nabla}
\newcommand{\hna}{\hat \nabla}
\newcommand{\hD}{\hat D}

\newcommand{\bphi}{{\bar \phi}}
\newcommand{\balpha}{{\bar \alpha}}
\newcommand{\bbeta}{{\bar \beta}}
\newcommand{\bJ}{{\bar J}}
\newcommand{\bz}{{\bar z}}

\newcommand{\bo}{\bar o}
\newcommand{\bm}{\bar m}
\newcommand{\bi}{\bar i}
\newcommand{\bX}{\bar X}
\newcommand{\bY}{\bar Y}
\newcommand{\bsi}{\bar \sigma}
\newcommand{\bla}{\bar \lambda}
\newcommand{\bga}{\bar \gamma}

\newcommand{\ep}{\epsilon}

\newcommand{\si}{\sigma}
\newcommand{\ga}{\gamma}

\begin{document}

\title{Positivity of the Bondi Energy in Einstein-Maxwell
Axion-dilaton Gravity}

\author{Marek Rogatko \thanks{
Supported in part by KBN grant 2 P03B 093 18.} \\
Technical University of Lublin \\
20-618 Lublin, Nadbystrzycka 40, Poland \\
rogat@tytan.umcs.lublin.pl \\
rogat@akropolis.pol.lublin.pl}

\date{\today}

\maketitle
\smallskip
{\bf PACS numbers:} 04.50.+h.\\
\begin{abstract}
We present a proof of the positivity of the Bondi energy
in Einstein-Maxwell axion dilaton gravity, being the low-energy
limit of the heterotic string theory. We consider the spacelike 
hypersurface which asymptotically approaches a null cone and on 
which the equations of the theory under consideration are given.
Next, we generalize the proof allowing the hypersurface 
having inner boundaries.
\end{abstract}


\baselineskip=18pt
\section{Introduction}
In the framework of general relativity, there are two distinct 
regimes where we can measure the mass. The first, at null infinity, 
yields the Bondi mass while the second, at spacial infinity, represents
the Arnowitt-Desser-Misner (ADM) mass. The main difference between
these two notions is that the Bondi momentum is dynamical while
the ADM one is not. The Bondi quantity is tightly bounded with an 
instant of retarded time and changes according to the radiation
which the source emits. The ADM four-momentum vector is fixed.
The tantalizing question is whether the physically reasonable 
gravitating systems can radiate more energy than they initially
have, i.e. this is a question whether the Bondi energy can become 
negative.
\par
During last years there were heavy attempts to prove the positivity 
of the Bondi mass. In a number of recent papers this long-standing
conjecture has been confirmed. Nester and Israel \cite{is} 
generalized the spinor method of Witten \cite{wi}
and Nester \cite{ne}
to present that a system satisfying the dominant
energy condition has a positive total energy and a timelike
four-momentum measured at spacelike or future null infinity.
\par
Horowitz {\it et al} \cite{hor}
proved that for a spacetime which was
asymptotically flat at null infinity and for which the dominant energy
condition held and existed a nonsingular spacelike hypersurface
which asymptotically approached the null cone of constant retarded time
near null infinity, the Bondi four-momentum was a future directed
timelike or null vector. 
The Bondi four-momentum vanished iff we had a flat spacetime in 
the domain of dependence for the considered hypersurface.
It was done by extending the Witten's
proof of the positivity of energy to the case of null infinity.
In \cite{ash} it was shown that neither the ADM nor the Bondi four-momentum of
an isolated system in general relativity can be null.
\par
The previous arguments presented by Schoen and Yau 
\cite{y1} were modified to
demonstrate the positivity of the Bondi mass \cite{y2}. For a given
spacelike hypersurface asymptotic to the null cone, they have 
constructed
a three-dimensional asymptotically flat initial data set which ADM
mass is not greater than the Bondi mass of the null cone under
consideration. The positivity of the Bondi mass follows from the
previously proved theorem 
\cite{y1}. Their method is very sensitive to the
choice of the spacelike hypersurface asymptotic to the null cone.
\\
Ludvigsen and Vickers \cite{lu} managed to prove positivity of the Bondi
momentum by means of choosing the hypersurface which consisted a
compact spacelike part and a null part in the asymptotic region and
considered the projection of the Witten's type of equation on it.
\par
Reula and Tod \cite{re}
gave a new type of the proof of the positivity of the
Bondi mass. They proved the existence of solutions of Witten's 
equation for
the case in which a spacelike slice on which the equation was given
was asymptotically null. They provided the unified treatment of the ADM
and Bondi mass. Among all,
by a suitable choice of boundary conditions the positivity of Bondi
energy was shown for the case where the hypersurface was no longer free
of inner boundaries, but was bounded by the union of finite number of
marginally trapped surfaces.
\\
Horowitz and Tod \cite{hor1}
considered asymptotically constant solutions to the
Weyl neutrino equation
on an asymptotically flat spacetime. They found that
there existed a collection of conserved vector fields
associated with these solutions, depended on the local 
stress energy of the
matter and which had the property that their integrals over spacelike
surfaces yielded the total ADM or Bondi four-momentum of the spacetime.
\par
Lately, in \cite{ss}
the new spinorial proof applying to both the ADM and
Bondi energies was achieved. In this attitude
no existence theorems on noncompact
hypersurfaces were needed to obtain the desire results.
\par
The main aim of the paper is to provide some continuity with our
previous works concerning  the problem of the ADM mass and the positive
mass theorem for black holes in the low-energy limit of heterotic 
string theory, in Einstein-Maxwell axion-dilaton  (EMAD) gravity.
We try to enlarge our considerations to the problem of the Bondi 
mass in EMAD system. The organisation of this paper is as follows.
In section 2 we review the basic filed equations of the theory 
under consideration and prove the positivity of the Bondi mass
in the case in which the spacelike hypersurface on which the 
equations are given is asymptotically null. We also provide the
generalization of this theorem allowing the hypersurface having 
inner boundaries. Section 3 summarizes our results.
\par
In our paper we shall use two-component spinor notation \cite{pen}. 
Spinor indices are denoted by capital letters. Our signature of 
$g_{\mu \nu}$ is $(+ - - -)$ and convention for the curvature tensor is
$2 \na_{[ \alpha } \na_{\beta ]} \eta_{\ga} = - R_{\alpha \beta \ga
\delta}\eta^{\delta}$. The Einstein equations are $G_{\mu \nu}
= - T_{\mu \nu}$.


\section{Positivity of the Bondi Mass }
The so-called low-energy limit of superstring theories provides
an interesting generalization of the Einstein-Maxwell (EM) theory. A
simplified model of this kind in an Einstein-Maxwell axion-dilaton
(EMAD) coupled system 
containing a metric $g_{\mu \nu}$, $U(1)$ vector fields
$A_{\mu}$, a dilaton $\phi$ and three-index antisymmetric tensor field
$H_{\alpha \beta \ga}$. The action has the form \cite{ka}
\be
I = \int d^4 x \sqrt{-g} 
\left [ - R + 2(\na \phi)^{2} + {1 \over 3} H_{\alpha \beta \ga}
H^{\alpha \beta \ga} -
e^{-2\phi} F_{\alpha \beta} F^{\alpha \beta} \right ] + I_{matter},
\label{ac}
\ee
where the strength of the gauge fields is descibed by
$F_{\mu \nu} = 2\na_{[\mu} A_{\nu]}$ and
the three index antisymmetric tensor is defined by the relation
\be
H_{\alpha \beta \ga} = \na_{\alpha}B_{\beta \ga} - A_{\alpha}F_{\beta
\ga} + cyclic .
\ee
In four dimensions $H_{\alpha \beta \ga}$ is equivalent to the
Peccei-Quin pseudoscalar and implies the following:
\be
H_{\alpha \beta \ga} = {1 \over 2}\ep_{\alpha \beta \ga \delta}
e^{4 \phi} \na^{\delta} a.
\label{def}
\ee
A straightforward consequence of the 
definition (\ref{def}) is that the action (\ref{ac}) can be written
as
\be
I = \int d^4 x \sqrt{-g} 
\left [	- R + 2(\na \phi)^{2} 
+ {1 \over 2} e^{4 \phi} (\na a)^2
- e^{-2\phi} 
F_{\alpha \beta} F^{\alpha \beta} - a F_{\mu \nu} \ast F_{\mu \nu}
\right ] + I_{matter},
\label{act}
\ee
where  $\ast F_{\mu \nu} = 
{1 \over 2} \ep_{\mu \nu \delta \rho} F^{\delta \rho}$ .
\par
Using the two-component spinor notation, we define a
Maxwell spinor by the relation
\be
F_{m n} = \phi_{MN}~ \ep_{M' N'} + \bphi _{M' N'}~ \ep_{M N},
\ee
where
\be
\phi_{A B} = {1 \over 2} F_{AB C'}{}{}^{C'},
\qquad \bphi_{A'B'} = {1 \over 2} F_{C}{}{}^{C}{}_{A'B'}.
\ee
Then equations of motion are given by
\ben
\na_{M}{}{}^{N'} \phi^{MN} &=& \na^{N}{}{}_{M'} \bphi^{M'N'} ,\\
\na_{M}{}{}^{N'} \left (
z \phi^{MN} \right ) - \na^{N}{}{}_{M'} \left (
\bz \bphi^{M'N'} \right )
&=& \cJ^{N' N}(matter) ,\\
\na_{\mu} \na^{\mu} \phi - {1 \over 2 } e^{4 \phi} (\na a)^2 - e^{2 \phi}
\left ( \phi_{MN} \phi^{MN} + \bphi_{M'N'} \bphi^{M'N'} \right ) &=& 0, \\
\na_{\mu} \na^{\mu} a + 4 \na_{\mu} \phi \na^{\mu} a + 2 i e^{- 4 \phi}
\left ( \bphi_{M'N'} \bphi^{M'N'} - \phi_{M N} \phi^{M N} \right ) &=& 0, \\
6 \Lambda \ep_{AB} \ep_{A'B'} + 2 \Phi_{ABA'B'} = T_{AA' BB'},
\een
where we have introduced a complex {\it axi-dilaton}
$z = a + i e^{-2 \phi}$, while $\Lambda = {R \over 24}$
and $\Phi_{ABA'B'}$ is the curvature spinor (sometimes called
Ricci spinor). The energy momentum tensor reads
\ben
T_{\mu \nu}(F, \phi, a) = 8 e^{-2 \phi} 
\phi_{M N}~ \bphi_{M' N'} &-& \ep_{M N}~ \ep_{M' N'}
\left [
2 (\na \phi )^2 + {1 \over 2} e^{4 \phi} ( \na a )^2 \right ] \\ \nonumber
&+&
4  \na_{M M'} \phi~ \na_{N N'} \phi 
+ e^{4 \phi}\na_{M M'} a~ \na_{N N'} a.
\een
To begin with we define
the supercovariant derivatives for two-component spinors
$(\alpha^{A}{}{}_{(i)}, \beta_{A' (i)})$. Their forms followed easily from 
the adequate definition in the case of four-component spinors
introduced in \cite{ro,ka}, namely
\be
\hna_{M M'}~ \alpha_{K (i)} = \na_{M M'}~ \alpha_{K (i)} +
{i \over 2} e^{2 \phi} \na_{M M'} a~ \alpha_{K (i)}
+ \sqrt{2} e^{-\phi} \phi_{K M}~ a_{(ij)}~ \beta_{M'}^{(j)},
\ee
\be
\hna_{M M'}~ \beta_{A' (i)} = \na_{M M'}~ \beta_{A' (i)} +
{i \over 2} e^{2 \phi} \na_{M M'} a ~\beta_{A' (i)}
- \sqrt{2} e^{-\phi} \bphi_{A' M'}~ a_{(ij)}~ \alpha_{M}^{(j)},
\ee
where $a_{(ij)}$ is the two-dimensional equivalent of the matrices 
introduced in \cite{cr}.\\
In what follows it will be convenient to define the following
quantities: 
\be
\delta \la_{A' (i)}^{(\alpha)} = \sqrt{2} \left (
\sqrt{2} \na_{A' B} \phi~ \alpha^{B}{}{}_{ (i)} +
{i \over \sqrt{2}}e^{2 \phi}  \na_{A'B} a~ \alpha^{B}{}{}_{(i)}
- 4 i e^{- \phi} \bphi_{ A'}{}{}^{C'}~ a_{(ij)}~ \beta_{C'}{}{}^{(j)}
\right ),
\label{l1}
\ee
\be
\delta \la_{A (i)}^{(\beta)} = \sqrt{2} \left (
\sqrt{2} \na_{A}{}{}^{B'} \phi~ \beta_{B' (i)} +
{i \over \sqrt{2}}e^{2 \phi}  \na_{A}{}{}^{B'} a~ \beta_{B' (i)}
+ 4 i e^{- \phi} \phi_{ A C}~ a_{(ij)}~ \alpha^{C (j)}
\right ).
\label{l2}
\ee
The motivation for introducing these quantities is twofold
(see e.g. \cite{sup}). First, 
we would like to achieve the desired mass bound and therefore the
specific factors appear in these definitions. 
Second, expressions (\ref{l1} - \ref{l2})  have motivation as 
the supergravity transformation laws of the appropriate particles 
in the associated supergravity model.
The EMAD gravity model constitutes the bosonic part of
$N = 4,~d = 4$ supergravity (the so-called $SU(4)$ version) with
one gauge field \cite{ka}.
\par
We shall consider first an everywhere spacelike, asymptotically null
slice $\Si$ (a surface without boundary) of an asymptotically flat
spacetime. 
Let $t^{a}$ be the unit normal to the hypersurface $\Si$,
let $h_{ab} = g_{ab} - t_{a}t_{b}$ be the metric induced on the
hypersurface $\Si$ and $D_{a} = h_{a}{}{}^{b} \na_{b}$,
where
$\na_{b}$ is the covariant derivative on $M$ with respect to the metric
$g_{ab}$ on $M$. Now it is easily seen that
the Witten's equations on the hypersurface $\Si$ 
may be written in the form
\ben
\hD_{AA'}~ \beta^{A'}{}{}_{(i)} &=& D_{AA'} ~\beta^{A'}{}{}_{(i)}
+ {i \over 2} e^{2 \phi} D_{AA'}a ~\beta^{A'}{}{}_{(i)}
+ \sqrt{2} e^{- \phi} t_{AA'} t^{MM'} \bphi^{A'}{}{}_{M'} ~a_{(ij)}~
 \alpha_{M}^{}{}^{(j)}, \\
\hD_{AA'}~ \alpha^{A}{}{}_{(i)} &=& D_{AA'} ~\alpha^{A}{}{}_{(i)}
+ {i \over 2} e^{2 \phi} D_{AA'}a ~\alpha^{A}{}{}_{(i)}
- \sqrt{2} e^{- \phi} t_{AA'} t^{MM'} \phi^{A}{}{}_{M} ~a_{(ij)}~
 \beta_{M'}{}{}^{(j)}.
\een
Let us introduce the future directed null vector
$\xi_{m}$ for arbitrary (smooth) spinor fields 
$(\alpha^{M}{}{}_{(i)}, \beta_{M' (i)})$
on an asymptotically flat spacetime as follows:
\be
\xi_{m} = \alpha_{M}{}{}^{(i)} \balpha_{M' (i)} 
+ \bbeta_{M}{}{}^{(i)} \beta_{M'(i)}.
\ee
Consider next,
the vector field $p_{a}$ defined by the relation \cite{hor1}
\be
p_{n} = \hna^{m} \hna_{[ n }\xi_{m ]}.
\label{pp}
\ee
From (\ref{pp}), one can deduce that $p_{n}$
is the divergence free, and that conserved
quantity obtained by integration over the hypersurface $\Si$
may be expressed as an integral over asymptotic two sphere $S$.
Consequently, we have
\be
\int_{\Si} d\Si_{a} p^{a} 
= \int_{S}d S^{n} t ^{m} \hna_{[ n} \xi_{m ]},
\ee
Taking into account that $\xi_{m}$ is divergence free due to the 
generalized Weyl equations, 
and using the definition of $\hD_{a}$ on a hypersurface $\Si$
with the unit normal vector $t^{a}$, i.e.
\be
\hna_{a} = \hD_{a} + t_{a}t^{b} \hna_{b},
\ee
one finds that
\ben
t^{AA'}~ \hna_{b} \alpha_{A (i)} \hna^{b} \balpha_{A'}{}{}^{(i)}
&=& t^{A A'} \hD_{b} \alpha_{A (i)} \hD^{b} \balpha_{A'}{}{}^{(i)}
+ 2 t^{AA'} \hD_{A' B} \alpha^{B}{}{}_{(i)} \hD_{A B'} 
\balpha^{B' (i)}, \\ \nonumber
t^{AA'}~ \hna_{b} \bbeta_{A (i)} \hna^{b} \beta_{A'}{}{}^{(i)}
&=& t^{A A'} \hD_{b} \bbeta_{A (i)} \hD^{b} \beta_{A'}{}{}^{(i)}
+ 2 t^{AA'} \hD_{A' B} \bbeta^{B}{}{}_{(i)} \hD_{A B'} 
\beta^{B' (i)}.
\een
After a fairly lenghty calculations one can show
\ben \label{le}
{1 \over 2} \int_{S} dS^{ab} \hna_{a} \xi_{b} &=& {1 \over 2} \int_{\Si}
d \Si~  T_{a b}(matter)~ t^{a} \xi^{b} + \\ \nonumber
&-& \int_{\Si} d \Si \left (
t^{A A'} \hD_{b} \alpha_{A (i)} \hD^{b} \balpha_{A'}{}{}^{(i)}
+ 2 t^{AA'} \hD_{A' B} \alpha^{B}{}{}_{(i)} \hD_{A B'} 
\balpha^{B' (i)}  \right ) \\ \nonumber
&-& \int_{\Si} d \Si \left (
t^{A A'} \hD_{b} \bbeta_{A (i)} \hD^{b} \beta_{A'}{}{}^{(i)}
+ 2 t^{AA'} \hD_{A' B} \bbeta^{B}{}{}_{(i)} \hD_{A B'} 
\beta^{B' (i)} \right ) \\ \nonumber
&+& 
\int_{\Si} d \Si~ \cK_{A A'} t^{AA'} \\ \nonumber
&+& \int_{\Si} d \Si \left [
\left ( \delta \la_{A'(i)}^{(\beta) } \right )^{\dagger}
\delta \la_{A }^{(\beta) (i)} +
\left ( \delta \la_{A (i)}^{(\alpha)} \right )^{\dagger}
\delta \la_{A'}^{(\alpha) (i)} 
\right ] t^{AA'} \\ \nonumber
&-& {1 \over \sqrt{2}} \int_{\Si} d \Si \left [
J_{m}(F, \phi) t^{m} (\balpha^{B'(i)} \beta_{B'(i)}) - 
\bJ_{m}(F, \phi) t^{m} (\alpha^{B (i)} \bbeta_{B (i)}) \right ].
\een
The energy momentum tensor $T_{ab}(matter)$ yields
\be
T_{ab}(matter) = T_{ab}(total) - T_{ab}(F, \phi, a),
\ee
while a complex current is defined as follows:
\be
J_{A A'}(F, \phi) = \na_{A' B} (e^{- \phi} \phi^{B}{}{}_{A}).
\ee
The quantity $\cK_{A A'}$ is given by
\ben \label{kk}
\cK_{A A'} &=& {i e^{2 \phi} \over 2}
\na_{A B'} \phi~ \na_{B A'} a~ \xi^{B B'} 
+ i e^{2 \phi}  \na_{A' M} \phi \na_{A M'} a \left (
\alpha^{M}{}{}_{(i)} \balpha^{M' (i)} - \bbeta^{M}{}{}_{(i)} 
\beta^{M' (i)}  \right ) \\ \nonumber
&+& 8 \sqrt{2} i e^{- \phi} \na_{A' B} \phi~ \phi_{C A}~ 
\bbeta^{( B (i)} \alpha^{C )}{}{}_{(i)}
+ 4 \sqrt{2} e^{\phi} \na_{A' B} a~ \phi_{A C}~
\bbeta^{ [ B (i)} \alpha^{C ]}{}{}_{(i)} \\ \nonumber
&+& complex \enskip conjugate.
\een
We assume that the matter energy-momentum tensor obeys the 
following condition:
\be
T_{ab}(matter)~ t^{a} \xi^{b} \ge \sqrt{2} \left [
J_{m}(F, \phi) t^{m} (\balpha^{B'(i)} \beta_{B'(i)}) - 
\bJ_{m}(F, \phi) t^{m} (\alpha^{B (i)} \bbeta_{B (i)})
\right ].
\label{en}
\ee
Condition (\ref{en}) is stronger than the dominant energy condition
ussually assummed in general relativity in order to prove the
positivity of the Bondi mass. This kind of requirement is also known in
Einstein-Maxwell gravity \cite{hul, mass} where it constitutes the key point
of the proof. Roughly speaking relation (\ref{en}) tells us that the
local energy density is greater or equal to the local densities of the
adequate charge densities.\\
Moreover we put additional conditions, namely
\be
\delta \la_{A (i)}^{(\alpha)} = \delta \la_{A' (i)}^{(\alpha)} =
\delta \la_{A (i)}^{(\beta)} = \delta \la_{A' (i)}^{(\beta)} = 0,
\label{c1}
\ee
and 
\be
\cK_{A A'} = 0.
\label{c2}
\ee
The relation (\ref{c1}) is motivated by the invariance of the entire
system under the supersymmetry transformations
written in the language of two-component spinors.
The other introduces 
relations amongs the fields in the considered theory.
Relations like (\ref{c2}) arise in various kinds of the extensions of 
Einstein-Maxwell theory to the case of the low-energy string theory,
both in proofs of the positivity of the ADM mass and the positivity
of black holes masses \cite{sup, ro}. 
\\
Thus one can see from the analysis of the right side of
the relation (\ref{le}) that taking into account (\ref{en}-\ref{c2})
the right-hand side is positive if the third and fifth 
term are equal to zero. We remark that the second 
and the fourth terms are automatically positive since the metric on 
the hypersurface $\Si$ is negative. However, the third and the fifth 
term are negative, so we equate them to zero, i.e.
one requires the following:
\be
\hD_{A' A} \alpha^{A}{}{}_{(i)} = 0, \qquad
\hD_{A' A} \bbeta^{A}{}{}_{(i)} = 0.
\label{c3}
\ee
There are elliptic first order equations on $\alpha^{A}{}{}_{(i)}$
and $\bbeta^{A}{}{}_{(i)}$.\\
Under the conditions (\ref{en}-\ref{c2}) one can ensure the positivity
of the right-hand side of equation (\ref{le}) if we find asymptotically 
constant solutions to (\ref{c3}) on the hypersurface $\Si$. 
In Einstein's gravity the existence theorem for such solutions was given 
by Tod and Reula in \cite{re}. \\
Following this method, for readers' convenience, we 
quote the main steps of the proof of the existence theorem
for (\ref{c3}). 
Now, let define the Hilbert space \cite{rr}
$\cH$ as the 
completion of  $C_{0}^{\infty}$ (smooth and compactly supported)
$\si^{A}{}{}{(i)}$
spinor fields on $\Si$, under the following norm:
\be
\left ( \parallel \si^{A}{}{}_{(i)}  \parallel_{\cH} \right )^2 =
\int_{\Si} d \Si~ \hD_{B A'} \si^{B}{}{}_{(i)}~ \hD_{A B'} 
\bsi^{B'}{}{}_{(i)}
 t^{A A'}.
\label{in}
\ee
Using the generalized Hardy lemma \cite{re} one finds
that elements of $\cH$ are measurable fields and if 
$\{ \si^{A}{}{}_{(i) m} \}$ are $C^{\infty}_{0}$ Cauchy 
sequences converging to
$\si^{A}{}{}_{(i)}$, then $\hD_{AA'} \left ( \si^{A}{}{}_{(i) m}
\right )$ weakly 
converges to the distributional derivative $\hD_{AA'} \si^{A}{}{}_{(i)}$
of $\si^{A}{}{}_{(i) m}$. 
The Hardy lemma in our case can be obtained by the method presented in 
\cite{re} when we change ordinary differential operator for
supercovariant one acting on the hypersurface $\Si$.
Thus, the integral (\ref{in}) makes sence 
for all elements of $\cH$. We shall seek the solutions of
(\ref{c3}) in the forms
\ben \label{ab1}
\alpha^{A}{}{}_{(i)} &=& \alpha^{A}_{\infty (i)} + a^{A}{}{}_{(i)},
\qquad a^{A}{}{}_{(i)} \in \cH, \\ \nonumber
\bbeta^{A}{}{}_{(i)} &=& \bbeta^{A}_{\infty (i)} + b^{A}{}{}_{(i)},
\qquad b^{A}{}{}_{(i)} \in \cH, 
\een
By means of the Riesz representation theorem (see e.g. \cite{rie})
one defines linear functionals
\ben \label{h1}
f_{1} \left ( \si^{A}{}{}_{(i)} \right ) &=&
- \int_{\Si} d \Si~ \hD_{A B'} \bsi^{B'}{}{}_{(i)} \hD_{B A'} 
\alpha^{B}_{\infty (i)} t^{AA'}, \\ \nonumber
f_{2} \left ( \si^{A}{}{}_{(i)} \right ) &=&
- \int_{\Si} d \Si~ \hD_{A B'} \bsi^{B'}{}{}_{(i)} \hD_{B A'} 
\bbeta^{B}_{\infty (i)} t^{AA'},
\een
Furthermore, using the Bondi coordinates \cite{boc} and the asymptotic
values of spin coefficients found by Exton {\it et al} \cite{ex},
having in mind the attitude of Bramson \cite{bra}, one can assert that
that $\alpha^{B}_{\infty (i)}$ and
$\bbeta^{B}_{\infty (i)}$ are asymptotically constant spinor fields.
Considering, next, the Cauchy-Schwartz inequality for spinors, one 
finds out
that $f_{1} \left ( \si^{A}{}{}_{(i)} \right )$
and $f_{2} \left ( \si^{A}{}{}_{(i)} \right )$ are bounded linear 
functionals on the Hilbert space. It leads to the
conclusion that there exist
such $a^{A}{}{}_{(i)}$ and $b^{A}{}{}_{(i)}$ that
\ben \label{h2}
f_{1} \left ( \si^{A}{}{}_{(i)} \right ) &=&
 \int_{\Si} d \Si~ \hD_{A B'} \bsi^{B'}{}{}_{(i)} \hD_{B A'} 
a^{B}{}{}_{(i)} t^{AA'}, \\ \nonumber
f_{2} \left ( \si^{A}{}{}_{(i)} \right ) &=&
 \int_{\Si} d \Si~ \hD_{A B'} \bsi^{B'}{}{}_{(i)} \hD_{B A'} 
b^{B}{}{}_{(i)} t^{AA'}.
\een
Now, applying (\ref{h1}) and (\ref{h2}), integrating by parts we obtain
\ben \label{rr}
\int_{\Si} d \Si~ \hD_{A'}{}^{ B} \hD_{B B'}
 \bsi^{B'}{}{}_{(i)} \left (
\alpha_{A \infty (i)} + a_{A (i)} \right )
t^{AA'} &=& 0, \\ \nonumber
\int_{\Si} d \Si~ \hD_{A'}{}^{ B} \hD_{B B'}
 \bsi^{B'}{}{}_{(i)} \left (
\bbeta^{A \infty (i)} + b_{A (i)} \right )
t^{AA'} &=& 0,
\een
for all $\si^{A}{}{}_{(i)} \in C^{\infty}_{0}$.
Thus, there exist $a^{A}{}{}_{(i)}$ and $b^{A}{}{}_{(i)}$ given by 
the relations (\ref{ab1}) satisfying the weak form of equations
\be
\hD_{A}{}{}^{B'} \hD_{B B'} \alpha^{A}{}{}_{(i)} =
\hD_{A}{}{}^{B'} \hD_{B B'} \bbeta^{A}{}{}_{(i)} = 0.
\label{pdif}
\ee
If $\alpha^{A}{}{}_{(i)}$ and $\bbeta^{A}{}{}_{(i)}$
are smooth then they satisfy the above equation in its strong form.
Now, we turn to the question whether $\alpha^{A}{}{}_{(i)}$ and 
$\bbeta^{A}{}{}_{(i)}$ satisfy the first order elliptic equations
(\ref{c3}). It can be shown by supposing a contradiction \cite{re}
, i.e. we assume
there are such $\bla_{B'} = \hD_{BB'} \alpha^{B}{}{}_{(i)} \ne 0$
and $\bga_{B'} = \hD_{BB'} \bbeta^{B}{}{}_{(i)} \ne 0$.
Substituting them into relation (\ref{le}) we obtain the contradiction.
All these conclude the proof of the existence theorem.

\vspace{0.25cm}
\par
In order to study boundary conditions for $\alpha^{A}{}{}_{(i)}$
and $\bbeta_{A (i)}$ spinors we fix an asymptotically null hypersurface
and choose a system of Bondi coordinates such that $\Si$ 
asymptotically approaches a null cone \cite{boc}. For a spinor basis 
$(o^{A}, i^{A})$ we introduce a null tetrad $l^{a}, n^{a}, m^{a}, 
\bm^{a}$ in such a way that $n^{a} = i^{A} \bi^{A'}$ and $l^{a} =
o^{A} \bo^{A'}$ be the ingoing and outgoing null vectors 
orthogonal to $\Si$ and $l_{a} n^{a} = 1$. The conjugate 
complex null vectors $m^{a} = o^{A} \bi^{A'}$ are normalized 
as follows: $m^{a} \bm_{a} = - 1$ and $o_{A} i^{A} = 1$.
\par
We expand $\alpha^{A (i)}$ and $\bbeta_{A (i)}$ in terms of 
this basis, namely
\ben
\alpha^{A (i)} = X_{1}^{(i)} o^{A} + Y_{1}^{(i)} i^{A}, \label{a} \\
\bbeta^{A (i)} = X_{2}^{(i)} o^{A} + Y_{2}^{(i)} i^{A}. \label{b}
\label{ab}
\een
The two-surface bivector is $l^{[ a} n^{b ]} dS$, so 
the right-hand side of equation (\ref{le}) may be written as
\be
{1 \over 2} \int_{S} dS^{ab} \hna_{a} \xi_{b} =
{1 \over 2} \int_{S} dS \left (
n_{a} D \xi^{a} - l_{a} \Delta \xi^{a} \right ),
\label{rh}
\ee
where $D = l^{a} \na_{a}$ and $\Delta = n^{a} \na_{a}$.
The Weyl equations may be expressed in the form
\ben \label{we}
i_{A} D \alpha^{A}{}{}_{(i)} &+& {i \over 2} e^{2 \phi} i_{A}
D a~ \alpha^{A}{}{}_{(i)} - o_{A} \bde \alpha^{A}{}{}_{(i)}
- {i \over 2} e^{2 \phi} o_{A} \bde a~ \alpha^{A}{}{}_{(i)} = 0, \\
o_{A} \Delta \alpha^{A}{}{}_{(i)} &+& {i \over 2} e^{2 \phi} o_{A}
\Delta a~ \alpha^{A}{}{}_{(i)} - i_{A} \de \alpha^{A}{}{}_{(i)}
- {i \over 2}e^{2 \phi} i_{A} \de a~ \alpha^{A}{}{}_{(i)} = 0, \\
\bi_{A'} D \beta^{A'}{}{}_{(i)} &+& {i \over 2} e^{2 \phi} \bi_{A'}
D a~ \beta^{A'}{}{}_{(i)} - \bo_{A'} \de \beta^{A'}{}{}_{(i)}
- {i \over 2}e^{2 \phi} \bo_{A'} \de a~ \beta^{A'}{}{}_{(i)} = 0, \\
\bo_{A'} \Delta \beta^{A'}{}{}_{(i)} &+& {i \over 2} e^{2 \phi} \bo_{A'}
\Delta a~ \beta^{A'}{}{}_{(i)} - \bi_{A'} \bde \beta^{A'}{}{}_{(i)}
- {i \over 2}e^{2 \phi} \bi_{A'} \bde a~ \beta^{A'}{}{}_{(i)} = 0,
\een
where $\de = m^{a} \na_{a}$ and $\bde = \bm^{a} \na_{a}$.\\
By virtue of (\ref{rh}) and the above Weyl equations 
the boundary term on $S$ imply
\ben \label{bo}
{1 \over 2} \int_{S} dS^{ab} \hna_{a} \xi_{b} &=&
{1 \over 2} \int_{S} dS \left [
\rho X_{1} \bX_{1} + \mu Y_{1} \bY_{1} + 2 Re \bY_{1} \left (
\de X_{1} + \beta X_{1} \right ) \right ] \\ \nonumber
&+& {1 \over 2} \int_{S} dS \left [
\rho X_{2} \bX_{2} + \mu Y_{2} \bY_{2} + 2 Re \bY_{2} \left (
\de X_{2} + \beta X_{2} \right ) \right ] \\ \nonumber
&+& {i \over 2} \int_{S} dS  e^{2 \phi} \de a \left (
X_{1} \bY_{1} + X_{2} \bY_{2} \right )
+ complex \enskip conjugate \\ \nonumber
&+& \sqrt{2} \int_{S} dS \left [
e^{- \phi} \phi_{1} \left ( \balpha^{A' (i)} \beta_{A' (i)}
\right ) + 
e^{- \phi} \bphi_{1} \left ( \alpha^{A (i)} \bbeta_{A (i)}
\right ) \right ],
\een
where the spin coefficients are
$\rho =  o^{A} \bde o_{A}, \enskip \mu = i^{A} \de i_{A}, \enskip
\beta = i^{A} \de o_{A}$ and a complex scalar costructed from 
a Maxwell spinor $\phi_{A B}$ is defined as
\be
\phi_{1} = \phi_{A B}~ o^{A} i^{B}.
\ee
By the definition, $\rho = 0$ on a future apparent horizon.
We impose the following
boundary conditions on $S$ when it is a future apparent horizon \cite{mass}
\be
\alpha^{A (i)} o_{A} = \beta^{A'(i)} \bo_{A'} = 0.
\label{co1}
\ee
On the other hand, on 
a past apparent horizon $\mu = 0$ and
one can input the boundary conditions \cite{mass} of the form:
\be
i^{A} \alpha_{A (i)} = \bi^{A'} \beta_{A'(i)} = 0.
\label{co2}
\ee
If the condition (\ref{co1}) for a future apparent horizon or
(\ref{co2}) for a past apparent horizon are satisfied, then 
the whole 
surface term given by (\ref{bo}) disappears.\\
To prove the existence theorem in the case of inner boundaries of $\Si$
we proceed similary to the case described above. Namely, one proves
the existence of solutions of equations (\ref{pdif}) with the adequate
boundary conditions. First we shall consider the case when $S$ is a
future apparent horizon. From relation (\ref{co1}), using (\ref{a})
one finds that
\be
Y_{1}^{(i)} \mid_{S}~ = Y_{2}^{(i)} \mid_{S}~ = 0.
\label{ab1}
\ee
The additional boundary conditions are as follows:
\be
\hD^{1'}{}{}_{B}~ \alpha^{B}{}{}_{(i)} \mid_{S}~ =
\hD^{1'}{}{}_{B}~ \bbeta^{B}{}{}_{(i)} \mid_{S}~ = 0.
\label{ab2}
\ee
Proceeding as before we show the existence of $a^{A}{}{}_{(i)}$
and $b^{A}{}{}{(i)} \in \cH$ such that $\alpha^{A}{}{}_{(i)}$
and $\bbeta^{A}{}{}_{(i)}$ given by the relations (\ref{ab1})
satisfy (\ref{pdif}). Since $\alpha^{A}_{\infty (i)}$ and 
$\bbeta^{A}_{\infty (i)}$ are chosen to vanish in the neighbourhood
of $S$, then it turns out that $a^{A}{}{}_{(i)}$ and $b^{A}{}{}_{(i)}$
assert the boundary conditions (\ref{ab1}) and (\ref{ab2}).
It can be seen considering equation (\ref{rr}),
integrating them twice by parts and using boundary conditions 
(\ref{ab1}) and the fact that $\si^{A} \in C^{\infty}_{0} \cap \cH$,
and $\si^{0'}$ can take arbitrary values on $S$.\\
The extension of the proof to the case when $S$ is past apparent
horizon is straightforward. From (\ref{b}) and (\ref{co2})
one obtains the boundary conditions
\be
X_{1}^{(i)} \mid_{S}~ = X_{2}^{(i)} \mid_{S}~ = 0.
\label{bb1}
\ee
and the suplementary boundary conditions on a future apparent horizon
yields
\be
\hD^{0'}{}{}_{B}~ \alpha^{B}{}{}_{(i)} \mid_{S}~ =
\hD^{0'}{}{}_{B}~ \bbeta^{B}{}{}_{(i)} \mid_{S}~ = 0.
\ee
The rest of the proof is the same as in the case of the 
future apparent horizon.

\vspace{0.25cm}
\par
To complete our considerations we shall cosider the boundary terms
at infinity. If $\alpha^{A}{}{}_{(i)}$ and $\bbeta_{A (i)}$ approach
the supercovariantly constant spinor at infinity, then having in mind
the attitude presented in \cite{wi}
it follows that
\be
{1 \over 2} \int_{S^{\infty}} dS^{ab} \hna_{a} \xi_{b} =
{1 \over 4} P_{m} \xi^{m}_{\infty} + {1 \over \sqrt{2}}
Q_{(F-\phi)} Re \left ( \alpha_{\infty}^{A (i)} 
\bbeta_{\infty A (i)} \right ) +
{1 \over \sqrt{2}}
P_{(F-\phi)} Im \left ( \alpha_{\infty}^{ A (i)}
\bbeta_{\infty A (i)} \right ),
\label{mass}
\ee
where the {\it dilaton-electric} charge and {\it dilaton-magnetic}
charge are defined respectively as
\be
Q_{(F-\phi)} =  2 \int_{S^{\infty}} dS
e^{- \phi_{\infty}}~ Re~ \phi_{1}, \qquad
P_{(F-\phi)} =  2 \int_{S^{\infty}} dS
e^{- \phi_{\infty}}~ Im~ \phi_{1}.
\ee
Inspection of equation (\ref{le}) reveals that the Bondi mass
$M_{BONDI} = \sqrt{P_{m} P^{m}}$ must be positive for all
$ \alpha_{\infty}^{ A (i)}$ and $\bbeta_{\infty A (i)}$,
hence
\be
P_{m} P^{m} \ge Q_{(F-\phi)}^2  + P_{(F-\phi)}^2.
\ee

\section{Conclusions}
In summary, we have studied the positivity of the Bondi mass
in the low-energy string theory, the so-called EMAD gravity.
We considered a non-singular spacelike hypersurface which 
asymptotically approached a null cone. 
One assumes that the conditions for
energy-momentum tensor (\ref{en}) and the specific relations 
amongs the fields in the theory (\ref{kk}) are satisfied.
Moreover, the entire system is 
invariant under the supersymmetry transformations. Then the Bondi 
four-momentum is a future directed timelike vector. The 
square of the Bondi mass
is greater than or equal to the sum of the squares of the total
{\it dilaton-electric} or {\it dilaton-magnetic} charges.
\par
The generalization of the above mentioned proof to the
case when the hypersurface has inner boundaries
was also provided. We considered
the case of a future and past apparent horizons establishing
the adequate boundary conditions for two-component spinors 
on them. One concludes, 
that if the above mentioned conditions for energy momentum tensor
and fields in EMAD gravity are satisfied
then the square of the
Bondi mass of the spacetime containing black hole
was greater or equal to the sum of the squares of the total
{\it dilaton-electric} and {\it dilaton-magnetic} charges of the 
black hole.

\vspace{1cm}
\noindent
{\bf Acknowledgment}\\
I would like to thank the unknown referees for very useful comments.\\

\eject

\end{document}